%% file: paper.tex
\documentclass[twocolumn,showpacs,aps,prl,superscriptaddress,floatfix,preprintnumbers]{revtex4}

\usepackage{graphicx}
\usepackage{dcolumn}
\usepackage{amsmath}
\usepackage{epsfig}

\RequirePackage{xspace}
\usepackage{relsize}
\def\babar{\mbox{\slshape B\kern-0.1em{\smaller A}\kern-0.1em
    B\kern-0.1em{\smaller A\kern-0.2em R}}}
\def\CP                {\ensuremath{C\!P}\xspace}
\def\piz   {\ensuremath{\pi^0}\xspace}

\def\Kstarp  {\ensuremath{K^{*+}}\xspace}

\def\Dbar    {\kern 0.2em\overline{\kern -0.2em D}{}\xspace}
\def\Dz      {\ensuremath{D^0}\xspace}
\def\Dzb     {\ensuremath{\Dbar^0}\xspace}
\def\Dstarp  {\ensuremath{D^{*+}}\xspace}

\def\Y#1S{\ensuremath{\Upsilon{(#1S)}}\xspace}

\def\invfb   {\ensuremath{\mbox{\,fb}^{-1}}\xspace}
\newcommand{\mev}{\ensuremath{\mathrm{\,Me\kern -0.1em V}}\xspace}
\newcommand{\mevc}{\ensuremath{{\mathrm{\,Me\kern -0.1em V\!/}c}}\xspace}
\newcommand{\mevcc}{\ensuremath{{\mathrm{\,Me\kern -0.1em V\!/}c^2}}\xspace}
\newcommand{\gevc}{\ensuremath{{\mathrm{\,Ge\kern -0.1em V\!/}c}}\xspace}
\newcommand{\gevcc}{\ensuremath{{\mathrm{\,Ge\kern -0.1em V\!/}c^2}}\xspace}
\def\ps   {\ensuremath{\rm \,ps}\xspace}
\def\pep2{PEP-II}

\newcommand{\rsdecay}{\mbox{\ensuremath{\Dz \to K^{-} \pi^{+} \pi^{0}}}}
\newcommand{\wsdecay}{\mbox{\ensuremath{\Dz \to K^{+} \pi^{-} \pi^{0}}}}
\newcommand{\wsdecayb}{\mbox{\ensuremath{\Dzb \to K^{-} \pi^{+} \pi^{0}}}}
\newcommand{\dm}{\ensuremath{\Delta m}}
\newcommand{\mKpp}{\ensuremath{m_{K\pi\pi^{0}}}}
\newcommand{\tKpp}{\ensuremath{t}}
\newcommand{\rightsign}{RS}
\newcommand{\wrongsign}{WS}

\newcommand{\BABARPubYear}    {06}
\newcommand{\BABARPubNumber}  {049}
\newcommand{\SLACPubNumber} {12036}

\def\figurebox#1#2#3{%
    \def\arg{#3}%
    \ifx\arg\empty
    {\hfill\vbox{\hsize#2\hrule\hbox to #2{\vrule\hfill\vbox to #1{\hsize#2\vfill}\vrule}\hrule}\hfill}%
    \else
    {\hfill\epsfbox{#3}\hfill}%
    \fi}

\begin{document}

\preprint{\babar-PUB-\BABARPubYear/\BABARPubNumber, SLAC-PUB-\SLACPubNumber} 

\title{\boldmath Search for \Dz-\Dzb\ Mixing
and Branching-Ratio Measurement\\
in the Decay \wsdecay
}

\input authors_jun2006

\date{\today}

\begin{abstract}
We analyze 230.4\invfb of data collected with the \babar\ detector
at the \pep2 $e^+ e^-$ collider at SLAC to search for evidence of
\Dz-\Dzb mixing using regions of phase space in the decay \wsdecay.
We measure the time-integrated mixing rate
$R_M =$ (0.023 $\mbox{}^{\rm +0.018}_{\rm -0.014}$\,(stat.) $\pm$ 0.004\,(syst.))\%,
and $R_M <$ 0.054\% at the 95\% confidence level, assuming \CP\ invariance. 
The data are consistent
with no mixing at the 4.5\% confidence level.
We also measure the branching ratio for \wsdecay\ relative
to \rsdecay\ to be (0.214 $\pm$ 0.008\,(stat.) $\pm$ 0.008\,(syst.))\%.
\end{abstract}


%

\maketitle


Mixing of the strong eigenstates
$|\Dz\rangle$ and $|\Dzb\rangle$, involving transitions
of the charm quark to a down-type quark, is expected
to have a very small rate in the Standard Model (SM).
Accurate estimates of this rate must consider
long-distance effects~\cite{DMixTheory},
and typical theoretical values of the
time-integrated mixing rate are
$R_M \sim \mathcal{O}(10^{-6}\textrm{--}10^{-4})$.
The most stringent constraint to date is
$R_M <$ 0.040\% at the 95\% confidence level~\cite{Zhang:2006dp}.
Because SM $D$ mixing involves only the first two
quark generations to a very good approximation,
the mixing-amplitude scale is
set by flavor-$\textrm{SU}(3)$ breaking,
and \CP\ violation is undetectable~\cite{DMixTheory}.

We search for the process $|\Dz\rangle\to|\Dzb\rangle$
by analyzing the decay of a particle
known to be created as a $|\Dz\rangle$~\cite{conjugates}.
We reconstruct the
wrong-sign (\wrongsign) decay \wsdecay,
and we distinguish doubly Cabibbo-suppressed (DCS)
contributions from Cabibbo-favored (CF) mixed contributions
in the decay-time distribution.
Because mixing amplitudes are small,
the greatest sensitivity to mixing is found when the amplitude
for a particular DCS decay is comparably small.
We increase our overall sensitivity to mixing by selecting
regions of phase space (\textit{i.e.}, the Dalitz plot) where the relative
number of DCS decays to CF decays is small.  This technique cannot be
performed with the two-body decay $\Dz \to K^+ \pi^-$, and it has not
been used to date. While the ratio of DCS to CF decay rates depends
on position in the Dalitz plot, the mixing rate does not.
From inspection of the Dalitz plots,
we note that DCS decays proceed primarily through the resonance
$\Dz\to \Kstarp\pi^-$, while
CF decays proceed primarily through
$\Dz\to K^-\rho^+$~\cite{Kopp:2000gv}.

We present the first search for $D$ mixing in the decay
\wsdecay.  The analysis method introduced
increases experimental accessibility to interference between
DCS decay and mixing without a full phase-space
parameterization.  Such interference effects
can be used to search for new physics contributions
to \CP\ violation.

The two mass eigenstates
\begin{equation}
\label{eq:massstates}
 |D_{A,B}\rangle = p|\Dz\rangle \pm q|\Dzb\rangle
\end{equation}
generated by mixing dynamics have different
masses $(m_{A,B})$ and widths $(\Gamma_{A,B})$, and we
parameterize the mixing process with the quantities
\begin{equation}
x \equiv 2\frac{m_{B} - m_{A}}{\Gamma_{B} + \Gamma_{A}},\ \ \ \ \ %
y \equiv \frac{\Gamma_{B} - \Gamma_{A}}{\Gamma_{B} + \Gamma_{A}}%
\textrm{.}
\end{equation}
If \CP\ is not violated, then $|p/q|=1$.
For a nonleptonic multibody \wrongsign\ decay, the time-dependent decay
rate, $\Gamma_{\textrm{WS}}(t)$,
relative to a corresponding right-sign (\rightsign)
rate, $\Gamma_{\textrm{RS}}(t)$, is
approximated by~\cite{Blaylock:1995ay}
\begin{eqnarray}
\label{eq:tdratemult}
 & {\displaystyle\frac{\Gamma_{\textrm{WS}}(t)}{\Gamma_{\textrm{RS}}(t)} =%
   \tilde{R}_D + \alpha\tilde{y}'\sqrt{\tilde{R}_D}\,(\Gamma t)%
   + \frac{\tilde{x}'^2+\tilde{y}'^2}{4}(\Gamma t)^2 } & \\
 &  0 \leq \alpha \leq 1\textrm{.} & \nonumber
\end{eqnarray}
The tilde indicates quantities that have been integrated over any choice
of phase-space regions.  $\tilde{R}_D$ is the integrated DCS branching ratio,
$\tilde{y}' = y\cos\tilde{\delta} - x\sin\tilde{\delta}$ and
$\tilde{x}' = x\cos\tilde{\delta} + y\sin\tilde{\delta}$,
where $\tilde{\delta}$ is an integrated strong-phase difference
between the CF and the DCS decay amplitudes,
$\alpha$ is a suppression factor that accounts for
strong-phase variation over the regions, and $\Gamma$ is the
average width.  The
time-integrated mixing rate
$R_M = (\tilde{x}'^2+\tilde{y}'^2)/2 = (x^2+y^2)/2$ is independent of decay mode.

We search for \CP-violating effects by fitting to the \wsdecay\ and
\wsdecayb\ samples separately.    We consider \CP\ violation
in the interference between the DCS channel and mixing, parameterized
by an integrated \CP-violating--phase difference
$\tilde{\phi}$, as well as \CP\ violation
in mixing, parameterized by $|p/q|$.
We assume \CP\ invariance
in the DCS and CF decay rates.  The substitutions
\begin{eqnarray}
 & \displaystyle \alpha\tilde{y}' \to \left|p/q\right|^{\pm 1}%
 (\alpha\tilde{y}'\cos\tilde{\phi} \pm \beta\tilde{x}'\sin\tilde{\phi}) & \\
 & \displaystyle (x^2+y^2) \to \left|p/q\right|^{\pm 2}(x^2+y^2) &
\end{eqnarray}
are applied to Eq.~\ref{eq:tdratemult},
using $(+)$ for $\Gamma(\wsdecayb)/\Gamma(\rsdecay)$ and $(-)$ for the
charge-conjugate ratio.  The parameter $\beta$ is a suppression factor
that accounts for $\phi$ variation in the selected regions.

We use 230.4\invfb of data
collected with the \babar\ detector~\cite{Aubert:2001tu}
at the \pep2 $e^+ e^-$ collider at SLAC.
The production vertices of charged particles are measured with
a silicon-strip detector (SVT), and their momenta are measured
by the SVT and a drift chamber (DCH) in a 1.5\,T magnetic field.
Particle types are identified
using energy deposition measurements from the SVT and DCH along with
information from a Cherenkov-radiation detector. 
The energies of photons are measured by an electromagnetic calorimeter.
All selection criteria were finalized
before searching for evidence of mixing in the data.
Selection criteria were determined from both study of the \rightsign\ sample and
past experience with other charm samples~\cite{babarmix}.

\begin{figure}[!tb]
\begin{center}
\includegraphics[width=\linewidth]{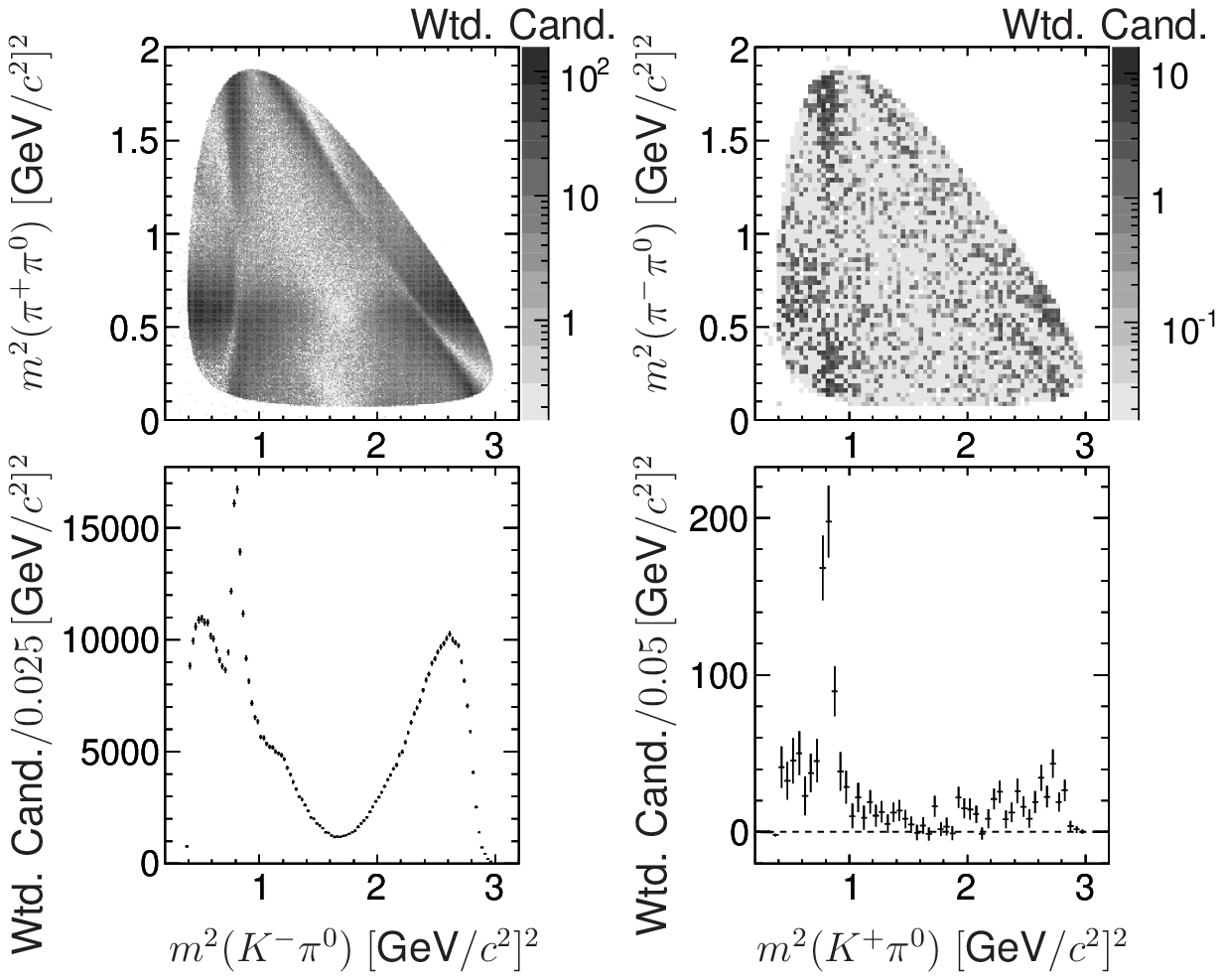}
\caption{
Dalitz plots and projections for \rightsign\ (left) and
\wrongsign\ (right) data.  An additional selection is made to
reduce peaking background in the events shown here, and
no $\sigma_t$ selection is made.
A statistical background subtraction~\cite{splot}
and a phase-space dependent efficiency correction have been applied
(\textit{i.e.}, candidates have been weighted). 
\vspace{-5ex}
}
\label{fig:dalitzplots}
\end{center}
\end{figure}

We reconstruct the decay $\Dstarp \to \Dz\pi_s^{+}$ and determine
the flavor of the \Dz\ candidate from the charge of the
low-momentum pion denoted by $\pi_s^{\pm}$.
We require $\pi_s^{\pm}$ candidates
to have momentum transverse to the beam axis $p_t > 120$\mevc.
We require \Dz\ candidates to have
center-of-mass momenta greater than 2.4\gevc, and 
the charged \Dz daughters must satisfy a likelihood-based
particle-identification selection. The identification
efficiency for both $K$ and $\pi$ is 90\%, and the misidentification
rate is 3\% (1\%) for $K$ ($\pi$) candidates.
We require photons from \piz\ decays to have a
laboratory energy $E_\gamma > 100\mev$,
and \piz\ candidates to have a laboratory momentum $p_{\piz} > 350\mevc$
and a mass-constrained--fit $\chi^2$ probability $> 0.01$.
The experimental width of the \piz-mass peak is
$\sigma_{m(\gamma\gamma)} \approx 6\mevcc$.
We accept candidates with an invariant mass
$1.74 < \mKpp < 1.98\gevcc$ and an invariant mass difference
$0.140 < \dm < 0.155\gevcc$, where
$\dm \equiv m_{K\pi\pi^{0}\pi_{s}} - \mKpp$.
We enhance contributions from
$\Dz \to K^- \rho^+$ and reduce
the ratio of DCS to CF decays
by excluding events
with two-body invariant masses in the ranges
$850 < m(K\pi^\pm,K\piz) < 950 \mevcc$.
Figure~\ref{fig:dalitzplots} shows the Dalitz plots for these decays.

The \Dstarp\ mass, \Dz\ mass, and \Dz\ decay time are derived from a
track-vertex fit~\cite{Hulsbergen:2005pu}.
A mass constraint is applied to the \piz\ candidate, and
the \Dstarp-decay vertex is constrained to the beamspot region, of size
$(\sigma_x,\sigma_y,\sigma_z) \approx (150\,\mu\textrm{m}, 10\,\mu\textrm{m}, 7\,\textrm{mm})$.
We select events for which the fit $\chi^2$ probability $> 0.01$.
From this fit, a \Dz\ decay time, \tKpp, and
uncertainty, $\sigma_t$, are calculated using the three-dimensional
flight path.  The full covariance matrix,
including correlations between the \Dstarp\ and \Dz\ vertices,
is used in the $\sigma_t$ estimate.
For signal events, the typical value of
$\sigma_t$ is near 0.23\ps.  We accept decays with
$\sigma_t < 0.5\ps$.  The \Dz\ lifetime is
(410.1 $\pm$ 1.5)\,fs~\cite{rppref}.

\begin{figure}[!tb]
\begin{center}
\includegraphics[width=\linewidth]{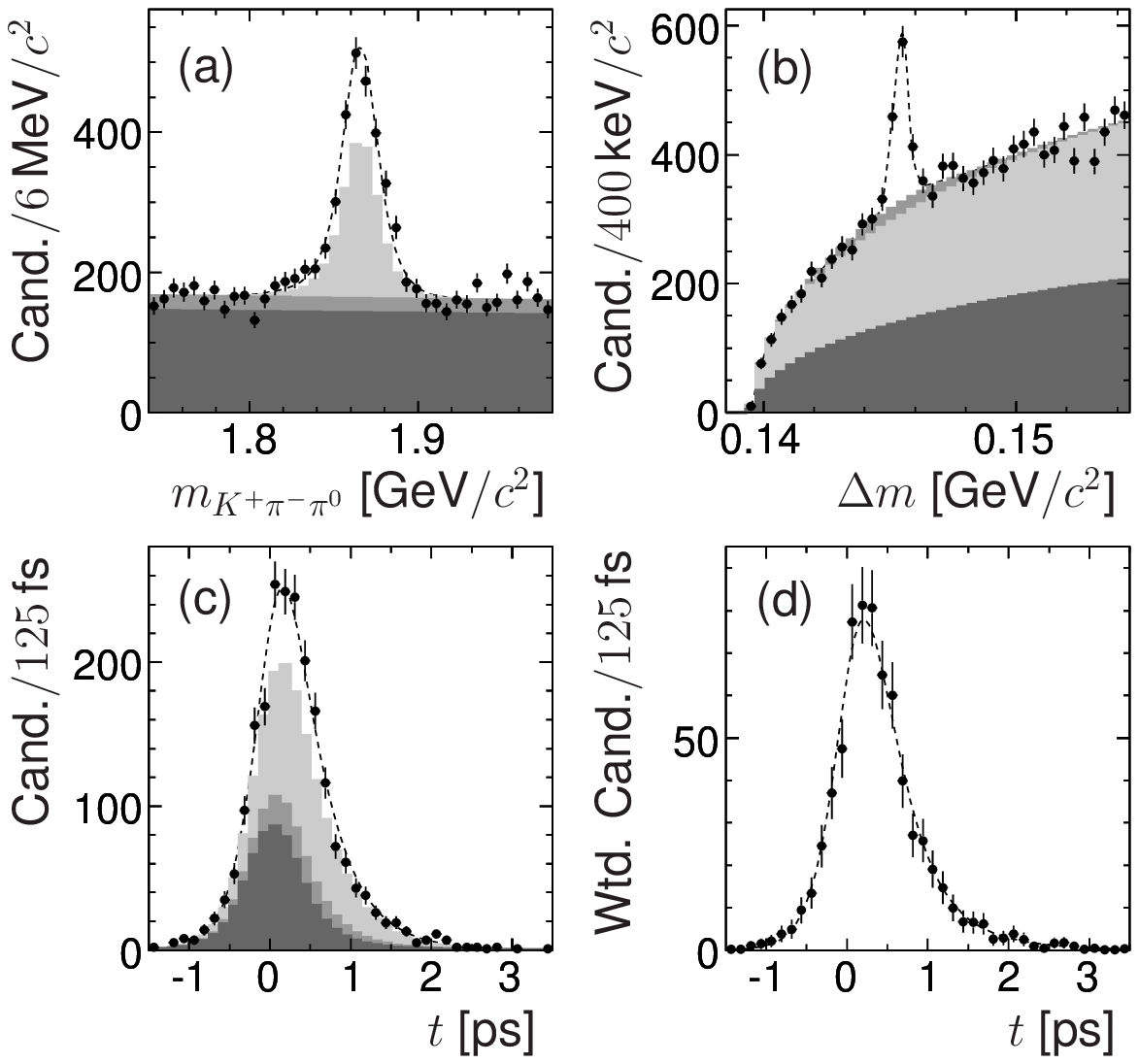}
\caption{
Distributions of \wrongsign\ data (points with error bars)
with fitted PDFs (dashed line) overlaid. The \mKpp\ distribution (a) requires
$0.1444 < \dm < 0.1464\gevcc$; the \dm\ distribution (b) requires
$1.85 < \mKpp < 1.88\gevcc$; and the \tKpp\ distribution (c)
requires both mass selections.
The data points in (d) show the \tKpp\ distribution
after applying a channel-likelihood signal projection~\cite{ChanLike,splot},
and the signal PDF is overlaid.
The error bars in (d) reflect Poissonian signal fluctuations only.
In (a)--(d), the white regions represent
signal events, the light gray misassociated
$\pi_s^{\pm}$ events, the medium gray
correctly associated $\pi_s^{\pm}$ with misreconstructed
\Dz\ events, and the dark gray remaining combinatorial background.
\vspace{-5ex}}
\label{fig:datafit}
\end{center}
\end{figure}

We first extract the signal
yields from a two-dimensional, unbinned, extended maximum
likelihood fit to the \mKpp\ and \dm\ distributions,
performed on the \rightsign\ and \wrongsign\
samples simultaneously.
The signal-shape parameters of the probability density function (PDF) describing
the \wrongsign\ sample are precisely determined by the large \rightsign\ sample,
and all associated systematic uncertainties are suppressed.
The width of the \dm\ peak is uncorrelated with
the width of the \mKpp\ peak, dominated by
\piz-momentum resolution, to first order.  However, there is
a second-order correlation in the signal between the two distributions.
Thus, the signal PDF has a width in \dm\ that
varies quadratically with \mKpp.
This feature significantly reduces the signal yield uncertainty.

Three background categories are included in the likelihood:
(1) correctly reconstructed \Dz\ candidates with a misassociated
$\pi_s^{+}$, (2) \Dstarp\ decays with 
a correctly associated $\pi_s^{+}$ and a misreconstructed \Dz,
and (3) remaining combinatorial backgrounds.
The first category has distributions in \mKpp\ and \tKpp\ 
of \rightsign\ signal decays and is distinguished using \dm.
The second category, peaking in \dm\ and distinguished
using \mKpp, has a \tKpp\ distribution similar to
\rightsign\ signal with a different characteristic lifetime.
The third category does not peak in either \mKpp\ or \dm\ and has a
\tKpp\ distribution empirically described by a Gaussian with a power-law tail.
Although the functional forms of the
background PDFs are motivated by simulations,
all shape parameters are obtained from a fit to the data.
The \mKpp\ and \dm\ projections of the two-dimensional
fit to the \wrongsign\ sample are shown in Fig.~\ref{fig:datafit}(a,b).

The signal yields from the fit to the $(\mKpp,\dm)$ plane are
listed in Table~\ref{tbl:candnumbers}.
Considering the entire allowed phase space,
and without the $\sigma_t$ selection,
we measure the branching ratio for \wsdecay\
relative to the decay \rsdecay\
to be (0.214 $\pm$ 0.008\,(stat.) $\pm$ 0.008\,(syst.))\%.
This result is consistent with previous measurements~\cite{expbr}
of this quantity and is significantly more precise.
For this measurement, a phase-space dependent efficiency correction
is applied to account for the different resonant populations in
CF and DCS decays.  The average efficiency of the \wrongsign\ sample
relative to the \rightsign\ samples is 97\%.
Phase-space dependent \piz\ selection efficiencies
dominate the systematic uncertainty.

\begin{table}[!tb]
\caption{
Signal-candidate yields determined by the
two-dimensional fit to the $(\mKpp,\dm)$
distributions
for the \wrongsign\
and \rightsign\ samples.  Yields
are shown (a) for the selected phase-space regions
used in
this analysis and (b) for the
entire allowed phase-space region.  Uncertainties
are those calculated from the fit, and no
efficiency corrections have been applied.}
\begin{center}
\begin{tabular}{llrr}
\hline
\multicolumn{2}{c}{\rule{0ex}{3ex}} &
\multicolumn{1}{c}{\Dz\ Cand.} &
\multicolumn{1}{c}{\Dzb\ Cand.} \\
\hline
 \rule{0em}{4.25ex}(\textit{a}) &
 \parbox{2em}{%
 \begin{tabular}{l}
 WS \\
 RS
 \end{tabular}} &
 \parbox{10em}{%
 \begin{tabular}{r}
 $(3.84 \pm 0.36) \times 10^2$ \\
 $(2.518 \pm 0.006) \times 10^5$
 \end{tabular}} &
 \parbox{10em}{%
 \begin{tabular}{r}
 $(3.79 \pm 0.36) \times 10^2$ \\
 $(2.512 \pm 0.006) \times 10^5$
 \end{tabular}} \\
 \rule[-2.75ex]{0em}{7.75ex}(\textit{b}) &
 \parbox{2em}{%
 \begin{tabular}{l}
 WS \\
 RS
 \end{tabular}} &
 \parbox{10em}{%
 \begin{tabular}{r}
 $(7.5 \pm 0.5) \times 10^2$ \\
 $(3.648 \pm 0.007) \times 10^5$
 \end{tabular}} &
 \parbox{10em}{%
 \begin{tabular}{r}
 $(8.1 \pm 0.5) \times 10^2$ \\
 $(3.646 \pm 0.007) \times 10^5$
 \end{tabular}} \\
 \hline
\end{tabular}
\label{tbl:candnumbers}
\vspace{-3ex}
\end{center}
\end{table}

The fitted shape parameters from \mKpp\ and \dm\
are used to determine the signal probability of each event
in a three-dimensional likelihood, $\mathcal{L}$, that is optimized
in a one-dimensional fit to $\tKpp$.
The \rightsign\ signal PDF in \tKpp\ is represented by an exponential function
convolved with a three-Gaussian detector-resolution function.
The Gaussians have a common mean, but different widths.  The width of each
Gaussian is a scale factor multiplied by $\sigma_t$, and $\sigma_t$
is determined for each event.
The three different scale factors, as well as the fraction of
events described by each Gaussian, are determined from
the fit to the data.  We find a $\Dz$ lifetime
consistent with the nominal value.

The WS PDF in $t$ is based on Eq.~\ref{eq:tdratemult} convolved with
the same resolution function as in the RS PDF.
The $\Dz$ lifetime
and resolution scale factors, determined by the fit to the
\rightsign\ \tKpp\ distribution,
are fixed.  We fit the \wrongsign\
PDF to the \tKpp\ distribution allowing yields and
background-shape parameters to vary.  The fit to the
\tKpp\ distribution is shown for the \wrongsign\ sample in
Fig.~\ref{fig:datafit}(c,d).

\begin{figure}[!tb]
\begin{center}
\includegraphics[width=\linewidth]{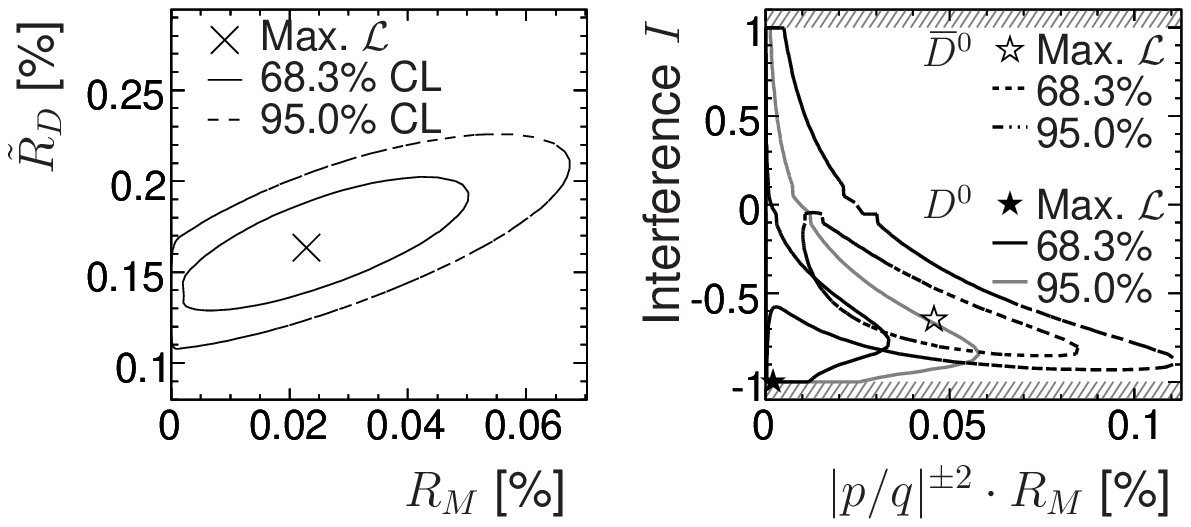}
\caption{
Contours of constant $\Delta\ln\mathcal{L}=1.15,3$, defining
68.3\% and 95.0\% confidence levels, respectively.  The contours on
the left are in terms of the integrated mixing rate, $R_M$, and
doubly Cabibbo-suppressed rate, $\tilde{R}_D$, assuming \CP\ invariance.
The contours on the right are in terms of $R_M$ and the normalized interference
$I = (\alpha\tilde{y}'\cos\tilde{\phi} \pm \beta\tilde{x}'\sin\tilde{\phi})/\sqrt{x^2+y^2}$,
for the \Dz\ and \Dzb\ samples separately.
On the left, the upward slope of the contour indicates
negative interference; on the right, the hatched regions are physically forbidden.
\vspace{-5ex}}
\label{fig:likelihood}
\end{center}
\end{figure}

The results of the decay-time fit, with and 
without the assumption of \CP\ conservation, are listed in
Table~\ref{tbl:results}.  The statistical uncertainty
of a particular parameter
is obtained by finding its extrema
for $\Delta\ln\mathcal{L}=0.5$.
Contours of constant $\Delta\ln\mathcal{L}=1.15,3$,
enclosing two-dimensional coverage probabilities of
68.3\% and 95.0\%, respectively,
are shown in Fig.~\ref{fig:likelihood}.
With a Bayesian interpretation of $\mathcal{L}$,
we find an upper limit $R_M <$ 0.054\% at the 95\% confidence level,
assuming \CP\ conservation.

In one dimension, $\Delta\ln\mathcal{L}$
changes its behavior near $R_M=0$ because
the interference term (the term linear in $t$ in Eq.~\ref{eq:tdratemult})
becomes unconstrained.  Therefore,
we estimate the consistency of the data with no mixing
using a frequentist method.  We generate 1000 simulated
data sets with no mixing but otherwise
according to the fitted PDF,
each with 58,800 events representing signal
and background in the quantities
\mKpp, \dm, and \tKpp.  We find 4.5\% of 
simulated data sets have a fitted value of $R_M$
greater than that observed in the data.
Thus, the observed data are consistent
with no mixing at the 4.5\% confidence level.

\begin{table}[!tb]
\caption{
Mixing results assuming \CP\ conservation
(\Dz\ and \Dzb\ samples are not separated) and
manifestly permitting \CP\ violation (\Dz\ and \Dzb\ samples are
fit separately).  The first listed uncertainty is
statistical, and the second is systematic.
Quantities that have been integrated over the selected
phase-space regions are indicated with tildes.
$\tilde{R}_D$ is not reported when allowing for
\CP\ violation because precise $\pi_s^{\pm}$ efficiency
asymmetries are unknown.}
\begin{center}
\begin{tabular}{cr|cr}
\hline
 & \multicolumn{1}{c}{\CP\ conserved} &
 & \multicolumn{1}{c}{\CP violation allowed} \\
\hline
 \rule{0em}{3ex}$R_M$ &
\multicolumn{1}{r}{$(0.023\ \mbox{}^{+0.018}_{-0.014} \pm 0.004) \%$} &
 &
\multicolumn{1}{r}{$(0.010\ \mbox{}^{+0.022}_{-0.007} \pm 0.003) \%$} \\
 \rule[-1.25ex]{0em}{1.25ex}$\tilde{R}_D$ &
\multicolumn{1}{r}{$(0.164\ \mbox{}^{+0.026}_{-0.022} \pm 0.012) \% $} &
 &
\multicolumn{1}{r}{} \\
\hline
 \rule{0em}{4.25ex}\rule[-3ex]{0em}{3ex}$\alpha\tilde{y}'$ &
 $-0.012\ \mbox{}^{+0.006}_{-0.008} \pm 0.002$ &
 \parbox{2em}{%
 \begin{tabular}{l}
 $\alpha\tilde{y}'\cos\tilde{\phi}$ \\
 \rule{0em}{2.75ex} \\
 \rule{0em}{3.25ex}$\beta\tilde{x}'\sin\tilde{\phi}$ \\
 \rule{0em}{2.75ex}
 \end{tabular}} &
 \parbox{10em}{%
 \begin{tabular}{r}
 \\
 \rule{0em}{2.75ex}$-0.012\ \mbox{}^{+0.006}_{-0.007} \pm 0.002$ \\
 \rule{0em}{3.25ex} \\
 \rule{0em}{2.75ex}$0.003\ \mbox{}^{+0.002}_{-0.005} \pm 0.000$
 \end{tabular}} \\
\hline
 \rule{0em}{3ex}& &
 \rule[-1.25ex]{0em}{1.25ex}$|p/q|$ & $2.2\ \mbox{}^{+1.9}_{-1.0} \pm 0.1$ \\
\hline
\end{tabular}
\label{tbl:results}
\vspace{-3ex}
\end{center}
\end{table}

We quantify systematic uncertainties by repeating the fits with
the following elements changed, in order of significance:
the background PDF shape in the \mKpp\ distribution,
the selection of events based on $\sigma_t$, the decay-time resolution
function, and the measured \Dz\ lifetime value.
Additionally, for $\tilde{R}_D$, we consider the absence of any Dalitz-plot
efficiency correction. The combined
systematic uncertainties are smaller than statistical uncertainties
by factors of 2--4.  The quantity
$\beta\tilde{x}'\sin\tilde{\phi}$, which quantifies a difference between
the \Dz\ and \Dzb\ samples, has a negligible systematic uncertainty
because positively correlated effects in the two samples cancel.

As a consistency check, we perform the decay-time fit to the entire
phase-space region populated by the decays \wsdecay.  The results
are consistent with Table~\ref{tbl:results}, with sensitivity
to $R_M$ preserved.  However, the interference term
obtained is different.  Figure~\ref{fig:likelihood}
indicates that both \Dz\ and \Dzb\ samples
prefer a large negative interference term
when the phase space is restricted
to suppress DCS contributions.  By contrast, 
when the interference term is integrated over
the entire Dalitz plot, it is found to be consistent
with zero, with uncertainties comparable to
those in this analysis.
The variation of the interference effect in different
phase-space regions motivates a detailed phase-space
analysis of this mode in the future.

In summary, we find that the data are consistent with the no-mixing hypothesis
at the 4.5\% confidence level, and we set an upper limit $R_M <$ 0.054\% at
the 95\% confidence level.
We measure the branching ratio for \wsdecay\
relative to \rsdecay\
to be (0.214 $\pm$ 0.008\,(stat.) $\pm$ 0.008\,(syst.))\%.

\input acknow_PRL.tex

\end{document}

%% file: authors_jun2006.tex
%
\author{B.~Aubert}
\author{R.~Barate}
\author{M.~Bona}
\author{D.~Boutigny}
\author{F.~Couderc}
\author{Y.~Karyotakis}
\author{J.~P.~Lees}
\author{V.~Poireau}
\author{V.~Tisserand}
\author{A.~Zghiche}
\affiliation{Laboratoire de Physique des Particules, F-74941 Annecy-le-Vieux, France }
% ...
\collaboration{The \babar\ Collaboration}
\noaffiliation

%% file: acknow_PRL.tex
We are grateful for the excellent luminosity and machine conditions
provided by our \pep2\ colleagues, 
and for the substantial dedicated effort from
the computing organizations that support \babar.
The collaborating institutions wish to thank 
SLAC for its support and kind hospitality. 
This work is supported by
DOE
and NSF (USA),
NSERC (Canada),
IHEP (China),
CEA and
CNRS-IN2P3
(France),
BMBF and DFG
(Germany),
INFN (Italy),
FOM (The Netherlands),
NFR (Norway),
MIST (Russia),
MEC (Spain), and
PPARC (United Kingdom). 
Individuals have received support from the
Marie Curie EIF (European Union) and
the A.~P.~Sloan Foundation.